\documentclass[english,12pt]{article}
\usepackage{color}
\usepackage{amssymb}
\usepackage{amsmath}
\usepackage{babel}
\usepackage{pifont}
\usepackage
[dvips,
colorlinks=true,
linkcolor=webgreen,%defined below
filecolor=webbrown,%defined below
citecolor=webgreen,%defined below
bookmarksopen=false,
]{hyperref}
\definecolor{webgreen}{rgb}{0,.5,0}
\definecolor{webbrown}{rgb}{.6,0,0}
\usepackage[T1]{fontenc}
\usepackage[cp1250]{inputenc}
\usepackage{array}
\usepackage[dvips]{graphicx}
\usepackage{amssymb}
\date{}
\definecolor{arcolor}{cmyk}{0.05,0.95,0.9,0.1}
\title{
Quantum Bargaining Games\\}
\author{Edward W. Piotrowski\\ Institute of Theoretical Physics,
University of Bia\l ystok,\\ Lipowa 41, Pl 15424 Bia\l ystok,
Poland\\ e-mail: \href{mailto:ep@alpha.uwb.edu.pl}{ep@alpha.uwb.edu.pl}\\
 Jan S\l adkowski\\ Institute of Physics, University of Silesia, \\ Uniwersytecka
4, Pl 40007 Katowice, Poland \\ e-mail:
\href{mailto:sladk@us.edu.pl}{sladk@us.edu.pl}
}
\begin{document}
\maketitle
\def\Z{{\bf Z\!\!Z}}
\def\R{{\bf I\!R}}
\def\N{{\bf I\!N}}
\begin{abstract}
We continue the analysis of quantum-like description of markets
and economics. The approach has roots in the recently developed
quantum game theory and quantum computing. The present paper is
devoted to quantum bargaining games which are a special class of
quantum market games without institutionalized clearinghouses.

\end{abstract}

 \vspace{5mm}

\section{Quantum bargaining} There have recently been important changes in the
paradigms of economics: economists discuss the role of the
Heisenberg uncertainty relation \cite{1} or even dare  claim that
quantum mechanics and mathematical economics are isomorphic
\cite{2}. These shocking changes have  probably been brought about
by the emergence of econophysics. Research on quantum computation
and quantum information allows to extend the scope game theory for
the quantum world \cite{3, 4}. Among various proposed qualitative
scientific methods only quantum theory does not allow to take no
account of the news phenomenon so persistent in social sciences
\cite{22}. Therefore the quantum-like description of market
phenomena has a remarkable chance of gaining favourable reception
from the experts. On the other hand only thorough investigation
may reveal if economics already is in or would ever enter the
domain of quantum theory. The present authors have given a general
description of quantum market games (q-games) in a recent paper,
\cite{5}. Among them one may distinguish the class of quantum
transactions (q-transactions) that is q-games without
institutionalized clearinghouses. This class comprises quantum
bargaining (q-bargaining) which are discussed in the present paper
and quantum auction (q-auction) to be discussed in a separate
paper. The participants of a q-bargaining game will be called
Alice ($A$) and Bob ($B$). We will suppose that they settle
beforehand who is the buyer (Alice) and who is the seller (Bob). A
two-way q-bargaining that is a q-bargaining when the last
condition is not fulfilled, will be analyzed in a separate paper.
Alice enter into negotiations with Bob to settle the price for the
transaction. Therefore the proper measuring apparatus consists of
the pair of traders in question. In q-auction the measuring
apparatus consists of a one side only, the initiator of the
auction.
\section{The Riemann sphere of polarization states} We will
identify  the space of Alice polarization states with the one
dimensional complex projective space $\mathbb{C}P^1$. Points in
$\mathbb{C}P^1$ will be called polarizations (or q-bits \cite{6}).
We will use the projective coordinates $\xi=(\xi_0,\xi_1)$. In
fact, we will introduce  a two dimensional Hilbert space
$\mathcal{H}_{\it{s}}$ and choose an orthonormal basis $(|{\mit
0}\rangle,|{\mit 1}\rangle)$, $|\xi\rangle=\xi_0 |{\mit 0}\rangle
+ \xi_1|{\mit 1}\rangle\in \mathcal{H}_{\it{s}}$. The scalar
product of two vectors $|\xi'\rangle,|\xi''\rangle
\in\mathcal{H}_{\it{s}}$ is given by
\begin{equation}
\label{ilskal}
\langle\xi'|\xi''\rangle={\bar{\xi}_{0}}'\xi_0''+\bar{\xi}_1'\xi_1''=
\bar{\xi}\cdot\xi \ .
\end{equation}
where $\bar{\xi}_k$, $k=0,1$ denotes the complex conjugate of do
$\xi_k$. The proportional vectors $|\xi\rangle$ and
 $t|\xi\rangle$
 ($t\in\mathbb{C}\setminus\{0\}$) will be identified. If we
 perform the transition from the homogeneous  coordinates to the
 non-homogeneous ones, $\underline{\xi}=(\xi_0,\xi_1)\rightarrow (1,z=\frac{\xi_1}{\xi_0})$,
$(0,\xi_1)\negthinspace\rightarrow\negthinspace\infty$, then the
projective space $\mathbb{C}P^1$ may be identified with the
compactified Gauss plain $\overline{\mathbb{C}}$. A point
$z\in\overline{\mathbb{C}}$ may in turn be identified with a point
in the two dimensional sphere $S^2$. This identification is given
by the bijection
$\mathcal{H}_{\it{s}}\negthinspace\rightarrow\negthinspace S^2$
\begin{equation}
\label{cayley} S^2\ni(x_1,x_2,x_3)=:\overrightarrow{r}(\xi)=
\frac{\langle\xi|\overrightarrow{\sigma}\xi\rangle}
{\langle\xi|\xi\rangle},
\end{equation}
where $\overrightarrow{\sigma}:=(\sigma_1,\sigma_2,\sigma_3)$ and
$\sigma_1$, $\sigma_2$ and $\sigma_3$ are the Pauli matrices. The
mapping $(\ref{cayley})$ is called Cayley-Klein parameterisation.
We will use all the three equivalent descriptions
$\mathbb{C}P^1\simeq \overline{\mathbb{C}}\simeq S^2$  of the
space of polarization states but one must remember that the
superposition (addition) of polarization states may only be
performed in the Hilbert space $\mathcal{H}_{\it{s}}$ \cite{6}.
The probability of measuring the strategy $|\xi''\rangle$ in the
state strategy $|\xi'\rangle$ is given by squared module of the
scalar product $(\ref{ilskal})$ of the states and may be expressed
in terms of the angle $\alpha$ between the corresponding unit
vectors $\overrightarrow{r}(\xi'),\overrightarrow{r}(\xi'')
\negthinspace\in\negthinspace \mathbb{R}^3$ (that is in $S^2$) by
the formula
$$\frac{|\langle\xi'|\xi''\rangle|^2}{\langle\xi'|\xi'\rangle\langle\xi''|\xi''\rangle}=\frac{1+
\overrightarrow{r}(\xi')\cdot\overrightarrow{r}(\xi'')}{2}=\cos^2\frac{\alpha}{2}.$$
Points in the Riemann sphere, $z\in\overline{\mathbb{C}}$, are in
the following 1-1 correspondence with  projective operators
$\mathrm{P}_{\overrightarrow{r}}$ acting on the Hilbert space
$\mathcal{H}_{\it{s}}$:
$$\mathrm{P}_{\overrightarrow{r}(\xi)}|\xi'\rangle:=
\frac{\langle\xi|\xi'\rangle}{\langle\xi|\xi\rangle}|\xi\rangle
.$$ They are usually written as
$\mathrm{P}_{\overrightarrow{r}(\xi)}=\frac{|\xi\rangle\langle\xi|}{\langle\xi|\xi\rangle}$.
In the orthonormal basis $(|{\mit 0}\rangle,|{\mit 1}\rangle)$\/
the projectors take the matrix form $P_{\overrightarrow{r}}$:
\begin{equation}
\label{stokes}
P_{\overrightarrow{r}}=\frac{I+\overrightarrow{r}\cdot\overrightarrow{\sigma}}{2},
\end{equation} where $I$\/ is the unit matrix. This representation
is referred to as the Stokes parametrization \cite{7} and is the
inverse one to the Cayley-Kleina parametrization
$(\ref{cayley})$\/. We will use the following interpretation of
the Alice polarization state
$|\xi\rangle_A\in\mathcal{H}_{\it{s}A}$  (that is of her
strategy). If she formulates the conditions of the transaction we
say she has the polarization ${\mit 1}$ (and is in the state
$|\overrightarrow{r}\rangle_A=|{\mit 1}\rangle$). In q-bargaining
this means that she put forward the price. In the opposite case,
when she decides if the transaction is made  or not, we say she
has the polarization $|{\mit 0}\rangle$\/. (She accepts or not the
conditions of the proposed transaction.) This is an analogy of the
isospin symmetry in nuclear physics which says that nucleon has
two polarization states: proton and neutron. The vectors $(|{\mit
0}\rangle,|{\mit 1}\rangle)$\/ form an orthonormal basis in
$\mathcal{H}_{\it{s}}$\/, the linear hull of possible Alice
polarization states. Bob's polarization is defined in an analogous
way.
\section{Polarizations of q-bargaining}
The states representing Alice and Bob strategies are fermions from
the physical point of view. This seemingly surprising statement
consist in noticing that the transaction in question is made only
if the traders have opposite polarizations (and even that is not a
guarantee of the accomplishment). Obviously, the states of Alice
and Bob became entangled if they enter into  q-bargaining. The
reduction of the state $|\xi\rangle_{\negthinspace A}$\/ (Alice)
to $|{\mit 1}\rangle_{\negthinspace A}$ or $|{\mit
0}\rangle_{\negthinspace A}$ always results in Bob winding up in
in the state $|{\mit 0}\rangle_{\negthinspace B}$\/ or $|{\mit
1}\rangle_{\negthinspace B}$, respectively. Therefore the space of
polarizations of  q-bargaining is isomorphic to the space of
polarizations of a single participant of the bargaining in
question ($\mathbb{C}P^1$). The homogeneous coordinates of
polarizations of  q-bargaining, that is a two dimensional complex
Hilbert space $\mathcal{H}_{\it{s}}\subset\mathcal{H}_{{\it
s}A}\bigotimes \mathcal{H}_{{\it s}B}$, is spanned by two
orthonormal vectors $|{\mit 10}\rangle:=
  |{\mit 1}\rangle_A|{\mit 0}\rangle_B$\/ and $|{\mit 01}\rangle:=
  |{\mit 0}\rangle_A|{\mit 1}\rangle_B$\/.
%Therefore q-bargainings as the forming them Alice and Bob
%polarizations are.
A market process resulting in q-bargaining is described by a
projection ${\mathrm{P}}_{|{\mit 1}\rangle}:\mathcal{H}_{{\it
s}A}\bigotimes\mathcal{H}_{{\it
  s}B}\rightarrow\mathcal{H}_{\it{s}}$\/. The result of the
projection depends on the choice of the basis
$(|{\mit0}\rangle,|{\mit1}\rangle)$\/ (it is sufficient to point
out the vector $|{\mit0}\rangle$\/). We will suppose that the
polarization of q-bargaining is determined by the pair of
participants in a unique way. \\

This polarization space has  a fascinating connection to
Lorentzian geometry (special relativity). We cannot help noting
here that, according to Penrose \cite{20,21}, $\mathbb{C}P^1\simeq
S^{2}$ may be identified with the heavenly sphere of light rays
coming to an observer. The group of conformal transformations of
the Riemannn sphere $PSL(2,\mathbb{C})$ is isomorphic to the
proper Lorentz group $SO_{0}(3,1)$ that preserves the Minkowski
metric. Penrose have also shown that the points on the heavenly
sphere  may be obtained from spinors so that any point on the
heavenly sphere corresponds to a proposition specifying the state
of a spinor ( in the space of propositions in the quantum logic
associated to the Jordan algebra
$h_{2}(\mathbb{C})$). \\

The arguments given in the following paragraph show that the state
$|{\mit 01}\rangle$ is more profitable for Alice than the state
$|{\mit 10}\rangle$\/. The skill of replication of the vector
$|{\mit 0}\rangle$\/ would allow Alice avoiding winding up in the
state $|{\mit 10}\rangle$\/. But quantum cloning is not possible
\cite{8} and this  makes the quantum bargaining nontrivial and
very interesting. It seems worth to notice that although the
polarization state cannot be cloned it may be transferred. We may
by using classical and quantum communication channels \cite{9, 10}
teleport Alice polarization state to another q-bargaining site. In
that way Alice may enter into negotiation with Bill instead of
Bob. Nevertheless, she will not be able to accomplish both
transaction simultaneously. This seems to be natural if one
recalls the common belief in undividity of attention. Alice
however may multiply her profits during a fixed interval by using
financial oscillators \cite{11}. To this end she must be able to
master  validity times of various financial instruments bought or
sold  in q-bargaining at different times. \\

We will say that Alice polarization $|\xi\rangle_{\negthinspace
A}$\/ {\em dominates} Bob' one $|\xi\rangle_{\negthinspace B}$\/ (
 $|\xi\rangle_{\negthinspace A}\stackrel{^{\hspace{.2em}{
 \mit{1}}}}{\smash{\succ}}|\xi\rangle_{\negthinspace
B}$) during q-bargaining with the polarization $|\mit 0\rangle$\/
if the probability of being in the state $|\mit 0\rangle$\/ is
greater for Alice than for Bob, that is
$$|\xi\rangle_{\negthinspace A}\stackrel{^{\hspace{.2em}{
\mit{1}}}}{\smash{\succ}}|\xi\rangle_{\negthinspace
B}\Leftrightarrow \frac{|\langle\mit0|\xi\rangle_{\negthinspace
A}|^2}{_{A}\negthinspace \langle\xi|\xi\rangle_{\negthinspace A}}>
\frac{|\langle\mit0|\xi\rangle_{\negthinspace
B}|^2}{_{B}\negthinspace \langle\xi|\xi\rangle_{\negthinspace B}}
\Leftrightarrow x_{3A}<x_{3B}.
 $$
This definition is justified by the asymmetry of profits made on a
deal by the same party having different polarizations
$|\mit0\rangle$\/ and $|\mit1\rangle$\/. More detailed analysis of
possible profits will be presented in the following paragraph. Of
course, if the polarizations are the same then the relation of
dominance is transitive. If we consider three parties $A$, $B$ and
$C$ and every pair of them enters into a q-bargaining then we have
$$ \bigl(|\xi\rangle_{\negthinspace A}\stackrel{^{\hspace{.2em}{
\mit{1}}}}{\smash{\succ}}|\xi\rangle_{\negthinspace B}\ and \
|\xi\rangle_{\negthinspace B}\stackrel{^{\hspace{.2em}{
\mit{1}}}}{\smash{\succ}}|\xi\rangle_{\negthinspace
C}\bigr)\Rightarrow |\xi\rangle_{\negthinspace
A}\stackrel{^{\hspace{.2em}{
\mit{1}}}}{\smash{\succ}}|\xi\rangle_{\negthinspace C}.
$$
In this case there is a statistical non-quantum model  \cite{7}
leading to results identical to those of the discussed here
q-bargaining. However, if the polarizations  are not constrained
by cultural customs or initial agreement  they usually are
different and it is possible that $$ |\xi\rangle_{\negthinspace
A}\stackrel{^{\hspace{.2em}{
\mit{1}}}}{\smash{\succ}}|\xi\rangle_{\negthinspace B}\ ,\
|\xi\rangle_{\negthinspace B}\stackrel{^{\hspace{.2em}{
\mit{\widehat{1}}}}}{\smash{\succ}}|\xi\rangle_{\negthinspace C}\
and \  |\xi\rangle_{\negthinspace C}\stackrel{^{\hspace{.2em}{
\mit{\widehat{\widehat{1}}}}}}{\smash{\succ}}|\xi\rangle_{\negthinspace
A}
$$
An extreme example is the situation when
$|\xi\rangle_A\sim|\mit0\rangle$\/, $|\xi\rangle_B\sim|\mit
\widehat{0}\rangle$\/ and $|\xi\rangle_C\sim|\mit
\widehat{\widehat{0}}\rangle$\/ already known from the popular
game {\it rock, paper and scissors} (RPS). If domination results
in profit asymmetry among participants (see below) then traders
buying or selling, say, shares issued by companies belonging to
Alice, Bob and Carol may be perceived  as if playing an RPS game.
A quantum version of the RPS game is discussed by Iqbal and Tool
\cite{12}. They showed that contrary to the classical case there
is a stable Nash equilibrium in the quantum RPS game. The observed
non-transitivity of dominance in bargaining is a case in point for
using quantum description of bargaining. Empirical verification
should decide if such approach is correct. Of course, one may tray
to realize q-bargaining games on quantum level where both parties
may be formed by coalitions and the appropriate states would be
superpositions of states of the members but this is beyond the
scope of the present analysis.
\section{Rationality of decisions of making bargain } Let us
recall that the Alice strategy is, besides its polarization
$|\xi\rangle_A\negthinspace\in
\negthinspace\mathcal{H}_{\it{s}A}$, given by the supply-demand
factor $|\psi\rangle_A \in\mathcal{H}_A$\/ \cite{5}. The total
state of q-bargaining,
$|\Psi\rangle=|{\xi}\rangle|\psi\rangle_A|\psi\rangle_B$\/,
belongs to the  tensor product of Hilbert spaces
$\mathcal{H}_{\it{s}}\bigotimes\mathcal{H}_{A}\bigotimes\mathcal{H}_{B}$\/.
Let us consider the following situation. Alice is going to buy
some commodity at the price $c$. The random variable $q:=-\ln c$\/
describes the possible profit if the transaction would come true.
To be more precise, the additive logarithmic return ratio
$\ln\frac{c_0}{c}=\ln c_0 - \ln c$\/ measures profits made by
Alice if the commodity bought at the price $c$ has a real value
$c_0$\/ to her. If on the other side Bob want to sell the
commodity at the price $c$ his potential profits describes the
random variable $p:=\ln c$\/. Note that that the supply of an
asset at the price $c$ may be perceived as demand of money for
which one pays a definite amount of the asset equal to $c^{-1}$\/.
Therefore there is no minus sign in the definition of the random
variable $p$. The axioms of quantum theory say that the
probability density of revealing Alice and Bob intentions
described by the random variables $q$ and $p$, respectively, is
given by
\begin{equation}
\label{eigenstosc} \frac{|\langle q|\psi\rangle_A|^2}{\phantom{}_A
\langle\psi|\psi\rangle_A}\, \frac{|\langle
p|\psi\rangle_B|^2}{\phantom{}_B \langle\psi|\psi\rangle_B}\;d q d
p\ ,
\end{equation}
where $\langle q|\psi\rangle_A$\/ is the probability amplitude of
offering the price $q$ by Alice who wants to buy and the demand
component of her state is given by
$|\psi\rangle_A\in\mathcal{H}_{A}$\/. Bob's amplitude $\langle
p|\psi\rangle_B$\/ is interpreted in an analogous way. Of course,
the "intentions" $q$ and $p$ not always result in the
accomplishment of the transaction. If Alice being in the
polarization state $|\mit 10\rangle$\/ offers $q$ that is the
price $c={\mathrm e}^{-q}$\/ then Bob accepts it only if $p\leq
-q$\/ (${\mathrm e}^{p} \leqslant {\mathrm e}^{-q}$\/). The
symmetry $(q,p,c)\rightarrow(p,q,c^{-1})$\/ dictates the
acceptability condition for Alice in the state $|\mit
01\rangle$\/. The transaction would be accomplished only if Alice
pays no more than she plans, $p\leq -q$\/. Therefore the
rationality of the decision does not depend on polarization and is
described by the condition
\begin{equation}
\label{dobicie} q+p\leq 0\ .
\end{equation}

If we change the convention so that the variable $q$ describes
profits of the trading with Alice party instead of hers then the
condition $(\ref{dobicie})$\/ takes the form $q+p\geq0$\/ with
accordingly changed interpretation of $p$. \\

Let us use the convenient Iverson notation \cite{14} in which
$[expression]$\/ denotes the logical value (1 or 0) of the
sentence $expression$. The logarithm of the transaction price,
$\ln c$\/ is the random variable $-q$ or $p$ depending on the
polarization of the transaction ($|\mit 10\rangle$\/ or $|\mit
01\rangle$\/, respectively) restricted by the condition
$(\ref{dobicie})$\/. Therefore to obtain the probability density
in the state $|\mit 10\rangle$\/ we have to integrate over $p$ the
density given by Eq.~$(\ref{eigenstosc})$\/ multiplied by
$[q+p\leq 0]$\/ (Iverson notation)
\begin{eqnarray}
%\begin{split}
&\int_{-\infty}^\infty[q+p\leq 0] \frac{|\langle
p|\psi\rangle_B|^2}{\phantom{}_B
\negthinspace\langle\psi|\psi\rangle_B}\;dp\;\frac{|\langle
q|\psi\rangle_A|^2}{\phantom{}_A
\negthinspace\langle\psi|\psi\rangle_A}\;dq\Bigr|_{q=-\ln c}=\\
&\int_{-\infty}^{\ln c} \frac{|\langle
p|\psi\rangle_B|^2}{\phantom{}_B
\negthinspace\langle\psi|\psi\rangle_B}\;dp\;\frac{|\langle -\ln
c|\psi\rangle_A|^2}{\phantom{}_A
\negthinspace\langle\psi|\psi\rangle_A}\;d\ln c
 \label{formprzet}
% \end{split}
\end{eqnarray}
In an analogous way one gets the distribution of the random
variable $\ln c$\/ for the polarization $|\mit01\rangle$
\begin{eqnarray}
%\begin{split}
&\int_{-\infty}^\infty[q+p\leq 0] \frac{|\langle
q|\psi\rangle_A|^2}{\phantom{}_A
\langle\psi|\psi\rangle_A}\;dq\;\frac{|\langle
p|\psi\rangle_B|^2}{\phantom{}_B
\langle\psi|\psi\rangle_B}\;dp\Bigr|_{p=\ln c}=\\
&\int_{-\infty}^{-\ln c} \frac{|\langle
q|\psi\rangle_A|^2}{\phantom{}_A \langle\psi|\psi\rangle_A}\;dq\;
\frac{|\langle\ln c|\psi\rangle_B|^2}{\phantom{}_B
\langle\psi|\psi\rangle_B}\;d\ln c
%\end{split}
\label{formprzet2}
\end{eqnarray}
Note that these distribution are not yet correctly normalized.

\section{The case when Alice bargains with the Rest of the World}
Let us now consider the possible influence of changes in the price
character of strategies of bargaining parties on their profits.
The analysis of Nash equilibrium states \cite{15} of participants
changing their polarizations via strategies being unitary
homographies on $\mathbb{C}P^1_A$ i $\mathbb{C}P^1_B$\/ will be
presented in a separate work.  To illustrate the course of
q-bargaining and the resulting profits let us consider a simple
model. Suppose that Alice state is given by a wave function being
a proper function of the demand operator \cite{5}. Then the
probability density $\frac{|\langle
q|\psi\rangle_A|^2}{\phantom{}_A
\negthinspace\langle\psi|\psi\rangle_A}$\/ , see
Eq.~$(\ref{formprzet})$\/ and $(\ref{formprzet2})$ is given by
Dirac delta function $\delta(q\negthinspace-\negthinspace a)$. We
interpret it as the strategy {\it I buy only if the price $c$
lower than ${\mathrm e}^{-a}$} (therefore we call the value
${\mathrm e}^{-a}$\/ the withdrawal price). Let now Carol
represent the Rest of the World ($RW$) strategy \cite{5} that is a
market on which the good Alice is going to buy fulfills the
demand-supply law (is not a {\it giffen} \cite{5}). Then,
according to the theorem discussed in Ref. \cite{5}, the
probability density $\frac{|\langle
p|\psi\rangle_{RW}|^2}{\phantom{}_{RW}
\negthinspace\langle\psi|\psi\rangle_{RW}}$\/ is normal
(gaussian). Choosing appropriate units of the good in question and
performing the scale transformation $p\rightarrow\sigma p$\/,
where $\sigma$ is the dispersion leads to the standard normal
distribution $\eta(p)$\/. The formulas $(\ref{formprzet})$\/ i
$(\ref{formprzet2})$\/ take the following form:
\begin{equation}
\label{formp} \int_{-\infty}^{\ln c}\eta(p)\,dp\;\delta(\ln c
+a)\;d\ln c
\end{equation}
and
\begin{equation}
\label{formp2} [\ln c+a\leq0]~\eta(\ln c)\;d\ln c \ .
\end{equation}
The integrals of $(\ref{formp})$\/ and $(\ref{formp2})$\/ over
$\ln c\in(-\infty,\infty)$ represent the probability of making
q-transaction $E([\ln c + a\leq0])$ (equal for both polarizations
$|\mit10\rangle$\/ and $|\mit01\rangle$). The time of waiting for
accomplishment of the transaction $\tau$ is a random variable with
a geometrical distribution \cite{13} and does not depend on
polarization. Its expectation value is equal to
\begin{eqnarray}
\label{qczas}
%\begin{split}
E(\tau)&=&\bigl(1+E([\ln c +a\leq0])\sum_{k=1}^\infty
k\,(1-E([\ln c +a\leq0]))^{k-1}\bigr)\,\theta\\
&=&\bigl(1 + (E([\ln c +a\leq0]))^{-1}\bigr)\,\theta \ ,
%\end{split}
\end{eqnarray}
where $\theta$\/ is the characteristic (mean) time of duration of
q-bargaining \cite{13}. Eq.~$(\ref{qczas})$\/ respects the time
Alice needs for selling the good (also equal to $\theta $). This
is a consequence of her strategy $\delta(q-a)$. The profit Alice
made in a sequence identical of q-bargaining is measured by the
profit intensity equal to \cite{15}
\begin{equation}
\label{aaaq} -\;\frac{p_{\mit10}E_{\mit10}(\ln
c)+(1-p_{\mit10})E_{\mit01}(\ln c)}{E(\tau)}\ ,
\end{equation}
where $p_{\mit10}:=|\langle\mit10|10\rangle|^2$\/ is the
probability of making the transaction with polarization
$|\mit10\rangle$\/. The distributions $(\ref{formp})$\/ and
$(\ref{formp2})$\/ are not normalized so the averaging
$E_{\mit10}()$\/ and $E_{\mit01}()$ should be performed with
appropriately normalized distributions $(\ref{formp})$\/ i
$(\ref{formp2})$\/, that is $\delta(\ln c +a)d\ln c$ and
$\frac{[\ln c+a\leq0]}{E([\ln c+a\leq0])}\eta(\ln c)d\ln c$\/. By
combing Eq.~$(\ref{aaaq})$ and Eq.~$(\ref{qczas})$\/ and
multiplying both the numerator and denominator by $E([\ln
c+a\leq0])$ we get the following function $\rho(a)$\/ giving the
Alice profit intensity (in units of $\theta $)
\begin{equation}
\label{aaaq2}
\rho(a)=\frac{\int_{-\infty}^{-a}\bigl(p_{\mit10}a-p_{\mit01}x\bigr)\eta(x)dx}
{1+\int_{-\infty}^{-a}\eta(x)dx}
\end{equation}
\begin{center}
 \begin{figure}[h]
 \label{strateg}
 \begin{center}
\includegraphics[height=6.25cm, width=6.25cm]{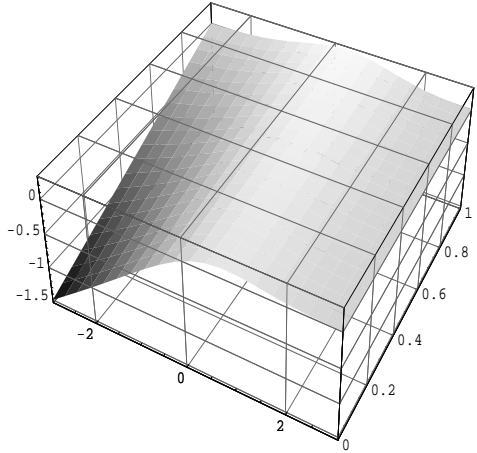}
\end{center}
\caption{Alice profit intensity in  $q$\/-bargaining with $RW$ as
a function of logarithm of withdrawal price $a\in[-2.5,2.5]$ and
probability
 $p_{\mit01}$\/ of  $q$\/-bargaining being in the state  $|\mit
01\rangle$\/.}
 \end{figure}
\end{center}

In the case when Rest of the World always decides to make the
transaction ($p_{10}=1$) the maximal profit intensity Alice may
get is equal $0.14028$\/ and corresponds to the withdrawal price
$a$ equal to $0.85096$\/. If on the other hand Rest of the World
always proposes the price then Alice get the maximal profit
intensity for the strategy $a=0.27063$\/. This value has the
property of being a fixed point of the function $\rho(a)$\/ ($\rho
$ is a contraction almost everywhere) \cite{13}. Therefore if
Alice does not know the parameters of the $RW$ distribution her
best method of achieving the condition $a=0.27063\sigma$ in a
polarization state $|\mit01\rangle$\/ is the simple iteration
$\rho(a)\negthinspace\rightarrow\negthinspace a$ leading to
self-correcting optimal algorithm  for getting the withdrawal
price $a$. In the discussed q-bargaining the polarization $|\mit
01\rangle$\/ is more profitable for Alice than $|\mit
01\rangle$\/. The characteristics of q-bargaining are presented in
Fig. 1 (diagram of the function $\rho=\rho(a,p_{\mit10})$). If
$p_{\mit 01}=1$\/ then any value of $a$  brings non-zero average
profits. On the other hand, if the q-bargaining has the
polarization $|\mit10\rangle$\/ (that is $p_{\mit 01}=0$\/) any
strategy with $a<0$\/ results in loses. Therefore we should not be
surprised that rivalry among sellers resulted in a market where
sellers display prices to give their potential clients better
bargaining positions.
\section{Temperatures of mixtures of q-bargaining} General
quantum bargaining involves mixed strategies and therefore should
be described in terms of probability density matrix (operator)
$\rho$ defined on the Hilbert space
$\mathcal{H}_{\it{s}}\bigotimes\mathcal{H}_{A}\bigotimes\mathcal{H}_{B}$\/.
Cases when the polarization of q-bargaining does not depend on
mutually independent mixed strategies of Alice and Bob should be
described by factorized density operators of the form
$\rho=\rho_{\it s}\rho_A\rho_B$\/, where
 $\rho_{\it s}$\/ is defined in $\mathcal{H}_{\it{s}}$,
 $\rho_A$\/ in $\mathcal{H}_A$\/, and $\rho_B$\/ in
 $\mathcal{H}_B$\/. The polarization state of q-bargaining,$\rho_{\it s}$\/, is now
a convex linear combination of pure states
$\mathrm{P}_{\overrightarrow{r}}$ belonging to $S^2$\/ given by
\begin{equation}
\label{wzorro} \rho_{\beta_{\it s},\overrightarrow{r}}=
\frac{{\mathrm e}^{\frac{\beta_{\it
s}}{2}\overrightarrow{r}\cdot\overrightarrow{\sigma}}}
{\mathrm{Tr}({\mathrm e}^{\frac{\beta_{\it
s}}{2}\overrightarrow{r}\cdot\overrightarrow{\sigma}})}=
\frac{1}{2}\Bigl(I+\overrightarrow{r}\cdot\overrightarrow{\sigma}\tanh\frac{\beta_{\it
s}}{2}\Bigr)\ ,
\end{equation}
where $\mathrm{Tr}$\/ denotes trace and $\beta_{\it
s}\in\mathbb{R}$ is the inverse of spin temperature of the system
described by the density operator $\rho_{\beta_{\it
s},\overrightarrow{r}}$\/.  The above formula may be simplified by
substitution $\frac{\beta_{\it s}}{2}\rightarrow\beta_{\it s}$\/
but our convention leads to formulas consistent with those
performed in the two-dimensional representation of the Lie algebra
$su(2)$. It follows from Eq.~$(\ref{wzorro})$\/ that the limits of
$\rho$\/ are projections $$ \lim_{\beta_{\it
s}\rightarrow\infty}\rho_{\beta_{\it s},\overrightarrow{r}}
=\lim_{\beta_{\it s}\rightarrow-\infty}\rho_{\beta_{\it
s},-\overrightarrow{r}} =P_{\overrightarrow{r}} \ .
$$ Let us suppose that $\beta_{\it s}\geq0$\/ in order to get
unamibiguous parameterization of the unit ball $K^2$\/ (apart from
the center). We may identify every polarization state
$\rho_{\beta_{\it s},\overrightarrow{r}}$ with some point $\rho\in
K_2$\/ . Strait lines joining those points with the center cut the
sphere $S_2$\/ at two antipodal points
$-\negthinspace\overrightarrow{r}$\/ and $\overrightarrow{r}$\/.
The proportion of the division of the segment
$\{-\negthinspace\overrightarrow{r},\overrightarrow{r}\}$\/ by
$\rho$ determines for every polarization state coefficients of its
representation as a convex combination of two projections
$$
\rho_{\beta_{\it
s},\overrightarrow{r}}=\frac{P_{\overrightarrow{r}}}{1+{\mathrm
e}^{-\beta_{\it s}}}+ \frac{P_{-\overrightarrow{r}}}{1+{\mathrm
e}^{\beta_{\it s}}}\ .
$$
This construction may be performed even if
$\beta\negthinspace=\negthinspace0$\/ when $\rho $ is the center
of $K^2$\/ but then the direction of $\overrightarrow{r}$\/ is
undetermined. There is another important quantity, namely Shannon
entropy $S$\/, that, instead of spin temperature, may be used to
parameterize polarization states. In our case Shannon entropy is
given by
$$
S(\beta_{\it s})=-\mathrm{Tr}(\rho_{\beta_{\it
s},\overrightarrow{r}} \ln\rho_{\beta_{\it
s},\overrightarrow{r}})= \frac{\ln\bigl(1+{\mathrm e}^{\beta_{\it
s}}\bigr)}{1+{\mathrm e}^{\beta_{\it
s}}}+\frac{\ln\bigl(1+{\mathrm e}^{-\beta_{\it
s}}\bigr)}{1+{\mathrm e}^{ -\beta_{\it s}}} \ .
$$
The present authors introduced in Ref. \cite{13} risk temperature,
being the inverse of Legendre multiplier $\beta$\/ . There is the
following relation between the dispersion $\sigma$\/ and the risk
temperature $\beta ^{-1}$ of the discussed above demand-supply
part of strategy of the Rest of the World (normal distribution)
$$
\sigma^2\,\tanh\frac{h_E\beta}{2\,\theta}={\mathrm{const.}}
$$
Here $h_E$\/ is the economic analogue of Planck constant. In
q-bargaining  the temperature of Alice strategy is zero because
her strategy may be perceived as the limit of the normal
distribution with dispersion tending to zero. This allows her to
perform arbitrage. The only remedy for limiting Alice profits in
trading with the Rest of the World consists in the lowering of the
$RW$ dispersion $\sigma$\/, that is the risk temperature
$\beta^{-1}$\/. As a result the optimal value of the parameter $a$
of Alice strategy will be shifted towards zero. Alice, to correct
her strategy, have to start a new iterative procedure to find the
new optimal value of $a$ and this will enlarge dispersion of her
strategy ($\beta_{Alice}^{-1}>0$). This process resembles heat
transfer from a hot thermostat ($RW$) to a cold one (Alice). As a
result the thermostat $RW$ cools down. Alice gets warmer and
subsequently while finding new value of $a$ lowers her temperature
(and entropy). \\

If we introduce a mechanism of coupling polarization of $RW$ with
its demand-supply part $\eta(p)$\/ being an integral Wigner
function with Gibbs weights \cite{5}, we would get a model
analogous to spin system interacting with phonon thermostat
\cite{16}. We envisage the investigation of q-bargaining with
methods characteristic to quantum quasi-equilibrium stochastic
processes. This should result in formulation of nonlinear
equations governing the dynamics of changes in risk and spin
temperatures. But investigation of q-bargaining without the
underlying thermostat $RW$ that dictates market prices seems to be
to abstract for application in economics. It is customary that the
polarization state in bargaining with the Rest of the World is
fixed in advance (e.g. sellers price). Therefore nontrivial
description of changes in spin temperature of q-bargaining may be
obtained only in models with at least three parties: A, B and $RW$
with A and B using different strategies with respect to themselves
than to $RW$. Such a minimal model is four-dimensional and
describes risk temperatures of A, B and $RW$ and spin temperature
of the
q-bargaining between A and B. \\

Note that spin temperature resembles another Legendre multiplier
connected to portfolio managing and introduced and discussed in
Ref. \cite{17}. Such coupled to the logarithmic rate of return
portfolio temperature (third kind of temperature of economic
process!) allows to compare skills of investors active in
different market or market conditions and not necessary during the
same intervals. For an aggressive market activity the logarithmic
rate of return cease to be additive and portfolio temperatures
acquire nonzero imaginary parts \cite{11}. The considered above
q-bargaining between Alice and $RW$ with both polarizations may be
perceived as multiple buying and selling of some financial asset
with normal distribution of  quotation of logarithm of its price.
So we are able to determine the portfolio temperature of Alice
strategy. Such thermodynamics of q-bargaining between Alice and
$RW$ would be presented elsewhere.

\section{Final remarks} If Bill intervene in q-bargaining between
Alice and Bob so that first transaction is made between Alice and
Bill and the second between Bob and Bill then we may call the
process  complex q-bargaining between Alice and Bob. The middlemen
(e.g. Bill) are filters from the quantum mechanical point of view.
In complex q-bargaining one may give up clearing of the
intermediate transactions. This would result in superposition of
amplitudes characterizing profits (logarithm of prices)  and
polarizations of complex q-bargaining. \\

The discussed connection between q-bargaining and condensed matter
physics suggest that q-bargaining might be performed with help of
quantum automata. If Penrose is right in his suggestions
concerning the process of thinking \cite{18} the phenomen of
q-bargaining might be possible to detect in present markets where
transactions are made due to non-computational algorithms. Does a
middleman performs the role of a "polarizator" facilitating making
transactions? \\

If we consider more realistic market of several assets then
q-bargaining will concern many price-like parameters
$c_1,c_2,\ldots$\/. So q-bargaining may be used for modeling
negotiations based on multi-criterion valuation of the offer. \\

Ancient Greeks, already 800 years BC, knew that the polarization
$|\mit01\rangle$\/ is more profitable for Alice, that is  she
accepts the transaction instead of proposing the price. Heziod
writes \cite{19} that Zeus accepting the method {\it one divides
the other chooses } in his deal with Prometheus , let the human to
divide. \\

\ \ \ {\bf Acknowledgments}. The authors would like to thank dr.
J. Eisert for stimulating and helpful discussions.
\def\urla{\href{http://econwpa.wustl.edu:8089/eps/get/papers/9904/9904004.html}{http://econwpa.wustl.edu:8089/eps/get/papers/9904/9904004.html}}
\def\urlb{\href{http://www.spbo.unibo.it/gopher/DSEC/370.pdf}{http://www.spbo.unibo.it/gopher/DSEC/370.pdf}}

\end{document}